\documentclass[aps,twocolumn,superscriptaddress,unsortedaddress,showpacs]{revtex4}
\usepackage{graphicx}
\usepackage{dcolumn}
\usepackage{bm}

\begin{document}

\title{Quantum blockade and loop current induced by a single lattice defect in graphene nanoribbons}
\author{Jie-Yun Yan}
\email{yan$\_$jieyun@iapcm.ac.cn} \affiliation{Institute of Applied
Physics and Computational Mathematics, P. O. Box 8009, Beijing
100088, China}
\author{Ping Zhang}
\affiliation{Institute of Applied Physics and Computational Mathematics, P. O. Box 8009,
Beijing 100088, China}
\author{Bo Sun}
\affiliation{Institute of Applied Physics and Computational Mathematics, P. O. Box 8009,
Beijing 100088, China}
\author{Hai-Zhou Lu}
\affiliation{Department of Physics, and Centre of Theoretical and
Computational Physics, The University of Hong Kong, Pokfulam Road,
Hong Kong, China}
\author{Zhigang Wang}
\affiliation{Institute of Applied Physics and Computational Mathematics, P. O. Box 8009,
Beijing 100088, China}
\author{Suqing Duan}
\affiliation{Institute of Applied Physics and Computational Mathematics, P. O. Box 8009,
Beijing 100088, China}
\author{Xian-Geng Zhao}
\affiliation{Institute of Applied Physics and Computational Mathematics, P. O. Box 8009,
Beijing 100088, China}

\begin{abstract}
We investigate theoretically the electronic transport properties in
narrow graphene ribbons with an adatom-induced defect. It is found
that the lowest conductance step of a metallic graphene nanoribbon
may develop a dip even down to zero at certain values of the Fermi
energy due to the defect. Accompanying the occurrence of the
conductance dip, a loop current develops around the defect. We show
how the properties of the conductance dip depend on the parameters
of the defect, such as the relative position and severity of the
defect as well as the width and edges of the graphene ribbons. In
particular, for metallic armchair-edges graphene nanoribbons,
whether the conductance dip appears or not, they can be controlled
by choosing the position of the single defect.
\end{abstract}

\pacs{73.63.-b, 72.10.Fk, 73.22.-f}
\maketitle

%73.63.-b Electronic transport in nanoscale materials and structures
%72.10.Fk Scattering by point defects, dislocations, surfaces, and
%         other imperfections ~including Kondo effect!
%73.22.-f Electronic structure of nanoscale
%         materials: clusters, nanoparticles, nanotubes, and nanocrystals
%75.75.+a Magnetic properties of nanostructures

\section{introduction}
Graphene has attracted intensive interest recently because of its
novel properties, such as the room-temperature quantum Hall effect
\cite{Zhang2005}
\cite{Novoselov2007} and the Dirac-equations-governed electrons \cite%
{Novoselov2005}\cite{Kane2005}. Moreover the properties of the
graphene nanoribbons\cite{CastroNeto2008} change notably due to
various edges \cite{Peres2006}, different symmetries
\cite{LiZY2008}, constrictions\cite{Rojas2006} and even controllable
defects by irradiating the material with electrons and ions
\cite{Ruffieux2000}\cite{Hashimoto2004}\cite{Ewquinazi2003}. Any
wanted structures could be achieved by engineering the graphene with
the help of state-of-art fabrication technologies, which supports
the graphene as a prospective candidate for significant applications
in electronics- and magnetics-related devices.

Defects have been revealed to be an influential factor in
determining the transport properties of structures in nanoscale. For
example, it has been known that defects can reduce or totally block
the quantum conductance in carbon nanotube
\cite{CNT1}\cite{CNT2}\cite{CNT3}. Graphene could be regarded as an
unrolled carbon nanotube, and hence the transport of graphene in the
presence of defects is also an interesting problem. Actually,
several kinds of defects in a single layer of graphene have already
been paid attention to, such as pentagon/heptagon defects
\cite{pentagon}\cite{LiuWM2008}, vacancies \cite{vacancy}, adatoms
\cite{adatom}\cite{adatom2}, substitution\cite{Peres2007},
disorder\cite{disorder} and even combinations of some of them
\cite{domains}.

Although the electronic energy band structures and magnetic
properties in the presence of defect have been investigated in both
graphene and graphene ribbons, how these properties manifest in
transport is still lacking in attention. Recently, it has been found
that the interior pentagon/heptagon defects\cite{LiuWM2008} or the
defects such as vacancies or disorder at the edges\cite{Li2008}
would lead to total quantum blockade in graphene nanoribbons at
certain Fermi energies. However, how the quantum blockade is
connected with the severity or the positions (not limited to the
edges) of the defect, and the size or the symmetry of the graphene
nanoribbons is not so clear. As the adatom defect would not change
the topological structure of graphene lattice to some extent, the
connection between these factors and the defect can be unambiguously
understood by comparing with the clean graphene system. As we will
show, all these factors are really critical on the influence of the
defect. Therefore, a careful study of the influence of the adatom
defect on the transport properties of graphene nanoribbons is
deserved.

In this paper, we investigate the electronic transport in narrow
graphene ribbons in the presence of a single defect, which could be
induced by the absorption of a hydrogen atom. The model of the
defect is based on the ab-initio calculation. We find that the
conductance profile is greatly reshaped by the defect. Compared with
the integer steps of perfect graphene nanoribbons, the conductance
in the presence of a single defect drops even down to zero in the
low energy regime, leading to a transformation from metal to
semiconductor. Accompanied by this global conductance blockade, a
loop current builds up around the defect. The conductances of the
defective graphene nanoribbons with different widths, edges, and
positions of the defect are studied. We find the conductance
blockade and the coexistence of the loop current sensitively depend
on the severity and relative position of the defect. For metallic
armchair-edges graphene nanoribbons, whether the conductance
blockade and the coexistence of the loop current develop or not,
they could be controlled by the position of the defect.

\section{model and results}
The transport properties in graphene ribbons under low bias voltages
are governed by the $\pi$ bonded electrons, which are modeled by the
tight-binding spinless Hamiltonian
\begin{equation}
\mathbf{H}=-t\sum_{\langle i,j\rangle }(c_{i}^{\dag }c_{j}+h.c.),
\label{Hamiltonian}
\end{equation}
where $c_{i}^{\dag}$ ($c_{i}$) creates (annihilates) an electron on
the site $i$, $t$ is the nearest-neighbor hopping energy and is set
as 2.7 $eV$ in this paper. The distance between the nearest neighbor
carbon atoms is $a$ ($\sim $1.42 {\AA}).

\begin{figure}[tbp]
\includegraphics[width=9cm]{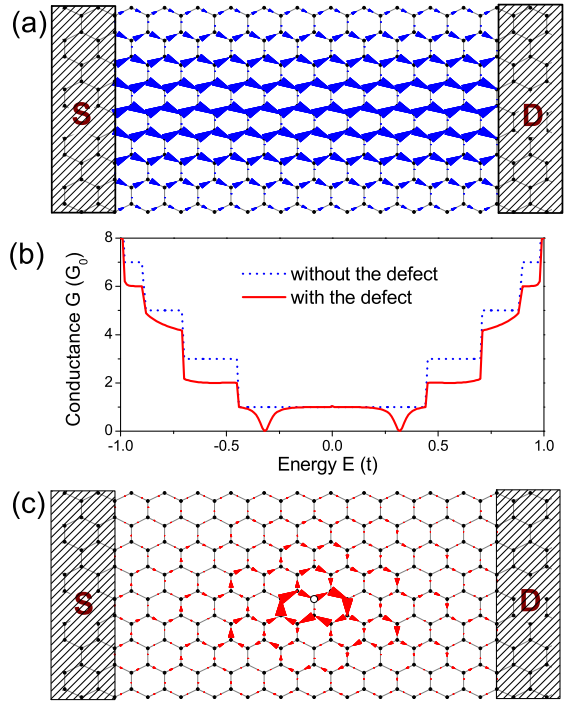}
\caption{(Color online) (a) The local electric current distribution
in the perfect graphene ribbons. The sizes of the arrow represent
the magnitude of the electric current between any two neighboring
sites, which are linearly normalized to the maximum value. The
source and drain are labeled with S and D respectively. (b) The
conductances of a perfect graphene nanoribbon (dot line) and the one
with a defect (solid line). The position of the defect is shown in
(c). (c) The local electric current distribution in the graphene
nanoribbon with a defect marked by a hollow circle. The Fermi energy
is $E=0.32t$.} \label{fig1}
\end{figure}

The conductance can be calculated by the Landauer-Buttiker formula
based on the Green function
method\cite{Datta1}\cite{LeeDH}\cite{ZhangJ}\cite{Anantram}. The
formalism separates the entire system into three parts, the central
region and two leads. To mimic the transport through infinitely long
ribbon with specific edge, two semi-infinite graphene leads are
connected to the central region, see Fig.\ref{fig1} (a). The zero
temperature conductance $G$ of the graphene nanoribbons is given by
$G(E)=\frac{2e^{2}}{\hbar }\mathtt{Tr}\left[ \mathbf{\Gamma}
_{S}(E)\mathbf{G}^{r}(E)\mathbf{\Gamma}
_{D}(E)\mathbf{G}^{a}(E)\right] $, where
$\mathbf{G}^{r(a)}(E)=(E\cdot\mathbf{I}-\mathbf{H}-\mathbf{\Sigma}
_{S}^{r(a)}(E)-\mathbf{\Sigma} _{D}^{r(a)})^{-1}$ is the retarded
(advanced) Green-function matrix of the central part,
$\mathbf{\Sigma} _{S(D)}^{r(a)}(E)$ is the retarded (advanced)
self-energy due to the source (drain) [labeled as S(D) in the
Fig.\ref{fig1} (a)], and $\mathbf{\Gamma} _{S(D)}(E)=i\left[
\mathbf{\Sigma} _{S(D)}^{r}(E)-\mathbf{\Sigma} _{S(D)}^{a}(E)\right]
$. Here $\mathbf{I}$ represents the unitary matrix. The local
current density at the Fermi level $E$ between two neighboring sites
$i$ and $j$ can be expressed as
\begin{equation}
i_{i\rightarrow j}(E)=\frac{4e}{\hbar }\mathtt{Im}[H_{ij}G_{ji}^{n}(E)],
\label{current}
\end{equation}%
where $\mathbf{G}^{n}=\mathbf{G}^{r}\mathbf{\Gamma}
_{S}\mathbf{G}^{a}$ is the electron correlation function and
$H_{ij}$ is the corresponding matrix element of the Hamiltonian.

In the absence of the defect, the well-known conductance of the
perfect graphene nanoribbon with zigzag edges is shown by the dot
line in Fig. 1(b). Due to the metallic nature of the system, there
exists no energy gap. The quantum conductance begins with $G_{0}$
(defined as $2e^{2}/\hbar $) and displays a typical step-increasing
feature as a function of the Fermi energy. We also show in Fig.
\ref{fig1} (a) the spatial distribution of the local current density
at a selective energy value of $E$=0.32$t$. One can see that the
current throughout the nanoribbon is uniform along the transport
direction and forms unambiguous paths. At this energy, the current
density amplitude at the central region of the nanoribbon is bigger
than that at the edge regions.

\begin{figure}[tbp]
\begin{center}
\includegraphics[width=6cm]{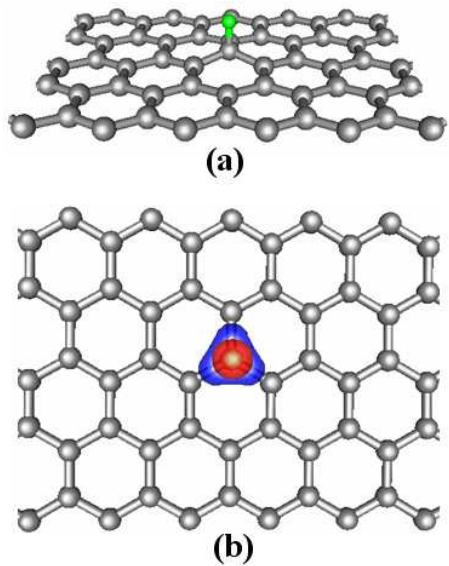}
\end{center}
\caption{(Color online) Adsorption of H on graphene: (a) The side
view of H/graphene system with an H adatom (the small green ball)
adsorbed at on-top site of the graphene (the big gray balls). (b)
The top view of the isosurface of the electron density difference
($\pm$0.05e/{\AA }$^3$). Charge flows from the blue into the red
regions.} \label{adatom}
\end{figure}

Now, let us focus on the effect of a single defect site on the
quantum conductance and local electric current distribution in the
graphene nanoribbons. To physically illustrate how to model the
presence of a single defect site by coupling a single carbon atom in
the graphene with an adatom atop, we have simulated the atomic
hydrogen adsorption on the graphene [see Fig. \ref{adatom}(a)] at a
low coverage of 0.02 monolayer (namely, one hydrogen atom presents
in a 5$\times $5 graphene supercell), by using density-functional
theory within the local density approximation and a supercell
approach. The total energy calculation shows that the most stable
adsorption site for H adatom is the on-top site and the H-C bond is
characterized by a hybridization of H 1$s$ and C 2$p_{z}$ states.
Prominently, it is found that the C atom beneath the H adatom is
pulled a little bit out of the garphene plane by H-C chemical
bonding, which obviously alters the coupling between this C atom and
its three neighboring C atoms and thus results in a single lattice
defect site. For further clarification, Fig. \ref{adatom}(b) plots
the calculated electron density difference $\Delta n$($\mathbf{r}$),
which is obtained by subtracting the electron densities of
noninteracting component systems, $n_{\text{graphene}}$
($\mathbf{r}$) +$n_{\text{H}}$($\mathbf{r}$), from the density
$n$($\mathbf{r}$) of the H/graphene system, while retaining the
atomic positions of the component systems at the same location as in
the H/graphene system. One can see that the charge redistribution
only occurs around the H-C bond and influences the three neighboring
C-C bonds, while the charge densities of other C atoms are not
affected at all.

Based on the physical picture revealed in Fig. \ref{adatom}, we
model the presence of a single defect lattice site by modifying the
interactions between this lattice site and its nearest neighboring
sites while keeping other hopping elements unchanged. Therefore we
change the hopping amplitude between the single defect site [hollow
circle in Fig. \ref{fig1}(c)] and its three nearest neighbors from
$t$ to, for example, $t_{d}$=0.3$t$. The conductance of the system
with the defect is calculated and plotted with the solid line in
Fig. \ref{fig1}(b). One can see that the conductance is prominently
reshaped by the presence of the single defect site. The explicit
changes include: (i) compared to the perfect case, the quantum step
feature disappears; (ii) the conductance drops for most Fermi
energies, (iii) zero-conductance dips appear at certain Fermi
energies. These defect-induced phenomena are understandable because
the defect is a kind of destructive factor for the electron
transmission in otherwise perfect graphene lattices. We now
concentrate on the conductance dip at $E$=0.32$t$. Since the
conductance drops to zero at this point, it means that the
transmission is totally blockaded by the defect. The dip also
implies the appearance of energy gap and consequently a
transformation from metal to semiconductor happens. For further
explanation to the conductance dip, we calculate the local current
density at the Fermi energy ($E$=0.32$t$) and plot the result in
Fig. \ref{fig1} (c). Compared to the ideal graphene nanoribbon
system [Fig. \ref{fig1}(a), plotted at the same energy $E$=0.32$t$],
it shows in Fig.\ref{fig1} (c) that the pattern of local current
density distribution is prominently changed in the defect-included
system and there appears loop current circling around the defect
site. The magnitude of this loop current around the defect can be
larger than the maximal current in Fig.\ref{fig1}(a). Note that the
current is normalized to the maximum value for each case. The loop
current would produce an effective magnetic field, which is
consistent with the previous experimental and theoretical
predication.\cite{vacancy}\cite{adatom}\cite{Yazyev2007}

The formation of the loop current could be explained as the
interference of all transport channels. In the two dimensional
graphene lattices, an electron could jump from source to drain via
many different transport channels. The local electric current
actually demonstrates the interference results of these channels. A
defect would cause the blockade of relative transport channels to
some extent and therefore change the local electric current. A
discussion using a simple one-dimensional model is given in
reference\cite{Li2008}. As the different transport channels are
numerous in the two dimensional graphene lattices, whether the loop
current appears and what kind of patterns it takes depend on the
whole environment of the defect, such as the edges and width of the
graphene ribbons, the position and severity of the defect, etc.
Therefore, we would dwell on the influence of these parameters on
the transport properties in the defective graphene ribbons.

\begin{figure}[tbp]
\begin{center}
\includegraphics[width=8cm]{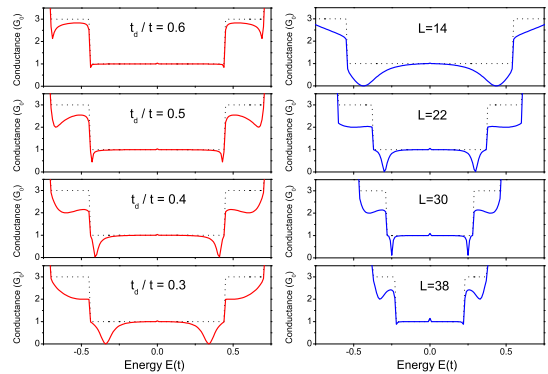}
\end{center}
\caption{(Color online) Left: Conductances with the ratio $t_d/t$
decreasing from top to bottom. The system is the same one as in
Fig.\ref{fig1}. Right: Conductances of the graphene nanoribbons with
different widths. In this case, the position of the defect and
length of the ribbon is the same as the one in Fig.\ref{fig1}. The
width changes by cutting or adding chains of C atoms at the bottom
of the graphene nanoribbons. L is number of the atoms connecting the
source lead. For every plot, the conductance of the perfect graphene
nanoribbon is shown for reference with the dotted lines.} \label{td}
\end{figure}

We can define the severity of the defect site by the ratio
$t_{d}/t$. The vacancy defect is actually the defect with
$t_{d}/t$=$0$. A question thereby arises: How effective could the
defect cause the conductance to drop to zero? In the following, we
will show that it depends on the competition between the width of
the graphene ribbon and the severity of the defect. For a narrow
graphene ribbon, the zero conductance exists in most cases when for
small ratios of $t_{d}/t$. In the left panel of Fig.\ref{td}, we
present the conductance of the same defective graphene ribbon as in
Fig.\ref{fig1} but with different $t_{d}/t$. We can see that as the
ratio $t_{d}/t$ decreases, the dip begins to appear in the lowest
conductance step and conductance reaches zero at $t_{d}/t$=$0.4$.
This critical value at which zero conductance is obtained is also
dependent on the width of the graphene ribbon. To show the influence
of the width, we modify the width of the defective graphene
nanoribbon shown in Fig.\ref{fig1} by cutting or adding chains of C
atoms on the lower edge of the ribbon. The width $L$ is denoted by
the number of chains. In the right panel of Fig.\ref{td}, the
conductances of the graphene ribbons with four different widths are
plotted. It is obvious that as the width of the graphene ribbon gets
wider, the dips become shallow and narrow and finally almost
disappear. For the case of $L=38$, the values of conductances almost
get equal between $E=0.22t$ and, for example, $E=0.1t$. However, the
loop current only exists near $E=0.22t$, not $E=0.1t$, although the
net current cannot be totally blocked by the defect.

\begin{figure}[tbp]
\begin{center}
\includegraphics[width=8cm]{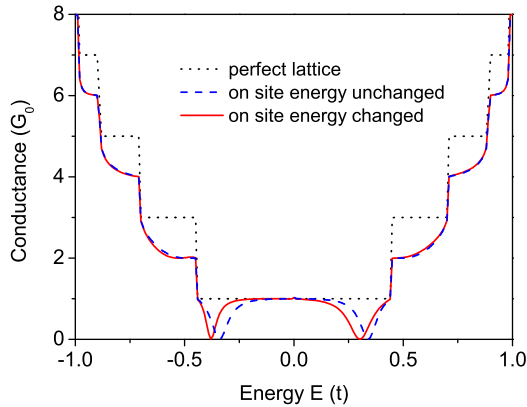}
\end{center}
\caption{(Color online) Conductance calculated for the case when the
on site energy of defect adds 0.1 $t$, shown with solid line. Other
parameters are the same as Fig.\ref{fig1}(b). For reference, the
case of a perfect lattice (dot line) and the one with a single
defect but assuming on site energy unchanged (dashed line) are also
shown.} \label{onsite}
\end{figure}
Besides the change of inter-site hopping matrix elements by the
coupling of a single lattice site with the adsorbed atom, the
on-site energy of this lattice site may also be changed due to the
bonding and anti-bonding hybridizations. The conductance by changing
the on-site energy of the defect site by 0.1$t$ is shown in
Fig.\ref{onsite}. We see that the conductance dip still appears but
lies at a shifted energy not far away. The distance depends on the
changed amplitude of on-site energy. At the conductance dip, the
loop current also exists without apparent difference from the one
shown in Fig.\ref{fig1} (c). So we conclude that the on-site energy
is not an influential factor to the occurrence of conductance dip
and formation of loop current induced by the single defect in this
model.
\begin{figure}[tbp]
\includegraphics[width=8cm]{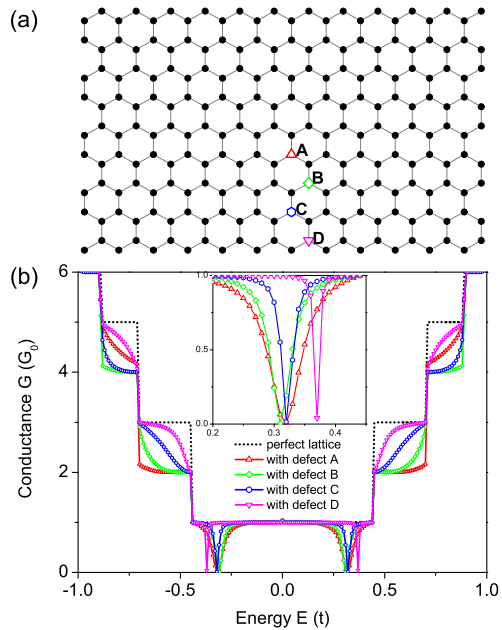}
\caption{(Color online) (a) The graphene nanoribbon with four
different defect sites to be considered respectively, which are
labeled with A, B, C and D. (b) Conductances of the graphene
nanoribbon with one defect located at one of these four positions
respectively. Inset is the amplification of the part where the
conductances drop to zero.} \label{fig2}
\end{figure}

We have also studied the influence of the position of the defect
site on the conductance of the graphene nanoribbons. It is found
that the horizontal shift of a single defect produces no changes to
the conductance, while the different defect positions in vertical
direction leads to different conductances and hence the center of
loop current patterns. In Fig. \ref{fig2}(a) we choose respectively
four sites (labeled by A, B, C and D) as a single defect site to
make a comparison. The conductance corresponding to these four
choices for the defect site are plotted in Fig. \ref{fig2}(b). One
can see that for each case the conductance drops to zero with a dip.
The width of the dip becomes smaller with the site going closer to
the boundary, see the inset in Fig.\ref{fig2} (b). By comparing the
current flowing through these sites in perfect lattice, we see that
the defect locating on the position with bigger current in the
perfect lattices would cause a conductance dip with a broader width.

\begin{figure}[tbp]
\includegraphics[width=8cm]{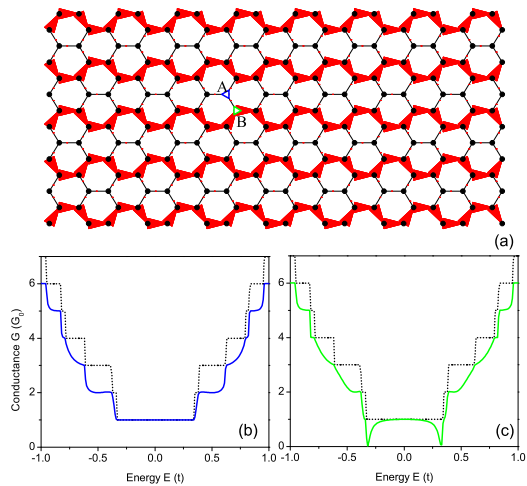}
\caption{(Color online) (a) Local electric current distribution
(arrows) in a metallic armchair-edges graphene ribbon. Sites A
(left-triangle) and B (right-triangle) represent two kind of defect
to be considered respectively. (b) Conductance of the system with
the defect located at position A. (c) Conductance of the system with
the defect located at position B. The conductance of the system
without any defect is also given with dotted lines for reference in
(b) and (c).} \label{fig3}
\end{figure}
The role of the presence of a single defect lattice site becomes
more important in the graphene nanoribbons with armchair edges. As
we know, a graphene nanoribbon with armchair edges has the zero
energy gap only when the number of atoms on a zigzag edge connecting
the source (or lead) is $3M-1$ ($M$ is an integer). This kind of
graphene nanoribbons hence has a quantum conductance beginning from
$G_{0}$ with increasing the electron Fermi energy. We plot the local
current distribution of the system without the defect in Fig.
\ref{fig3}(a), which clearly displays a series of well-separated
pathes for current flow. Each path has the same weight in current
amplitude. It is the very result of interference between the
transport channels in the metallic armchair-edges graphene
ribbons.\cite{Wakabayashi1999} According to this stripe current
distribution, the lattice sites in the graphene nanoribbon can be
therefore classified into two parts, i.e., sites like A [see Fig.
\ref{fig3}(a)] carrying current flow and sites like B without
carrying current flow. We set the single defect site to be on site A
or on site B, and then calculate the corresponding quantum
conductance respectively. The results are shown in Fig.
\ref{Hamiltonian}(b) and \ref{fig3}(c), respectively. Remarkably, it
is shown in Fig. \ref{fig3}(b) that when a single defect is placed
on site A, no conductance dip occurs at all and the first quantum
conductance step is invariant upon the presence of the defect.
Whereas, when the defect is placed on site B, two dips develop in
the conductance spectrum, as shown in Fig. \ref{fig3}(c). Again, at
the energy position $E$=0.32$t$ of the conductance dip, the spatial
distribution of the internal current (not shown here) is featured by
a loop current around the defect site. Thus, the lattice sites B are
more important than the lattice sites A for the transport in this
kind of armchair-edges graphene ribbons. This is an important
ingredient for practical control of the graphene transport by
manipulating the position of a single defect site. From the point of
the interference among different transport channels mentioned above,
the defect in site B gives a kind of destructive interference. The
local current density distribution pattern of the perfect metallic
armchair-graphene ribbons indicates that the total conductance is
mainly contributed by several parallel and independent channels. At
the site A, the local density is zero. In other words, the site A is
decoupled and isolated from these channels. That is why the graphene
ribbon is immune to the defect on site A.

\section{conclusion}
In summary, we have studied the influence of a single defect on the
electron transport properties in narrow graphene nanoribbons with
different edges, widths and positions of the defect. It has been
found that the conductances may drop even down to zero at certain
Fermi energies in the presence of a single defect site for metallic
graphene nanoribbons. The conductance dip is accompanied by the
formation of a remarkable loop current around the defect. The width
of the conductance dip is sensitive to the position as well as to
the severity of the defect site. For the graphene nanoribbons with
armchair edges, it has been found that the transport properties,
including the low-energy conductance step, the conductance dip and
the loop current, can be controlled by choosing the position of the
single defect. This result enlightens us on controlling the
conductance of the graphene nanoribbons by simply manipulating the
defects due to the absorbed atom.

%%%%%%%%%%%%% appendix %%%%%%%%%%%%%%

\end{document}